\documentclass[printer]{aa}
\usepackage{psfig}

\newcommand{\ark}{\hbox{Ark~564}}
\newcommand{\etal}{et al.}
\newcommand{\asca}{{\it ASCA}}
\newcommand{\xte}{{\it RXTE}}
\newcommand{\xmm}{{\it XMM-Newton}}
\newcommand{\rosat}{{\it ROSAT}}
\def\simlt{\lower.5ex\hbox{\ltsima}}            % < over ~
\def\simgt{\lower.5ex\hbox{\gtsima}}            % > over ~

\def\flux{erg\,cm$^{-2}$\,s$^{-1}$ }
\def\la{~\raise.5ex\hbox{$<$}\kern-.8em\lower 1mm\hbox{$\sim$}~}
\def\ma{~\raise.5ex\hbox{$>$}\kern-.8em\lower 1mm\hbox{$\sim$}~}

\begin{document}
\title{XMM$-$Newton observation of the NLS1 Galaxy  Ark~564: 
I.~Spectral analysis of the time-average spectrum}

\author{I.E. Papadakis\inst{1} \and W. Brinkmann\inst{2} 
\and M.J. Page\inst{3} \and I. M$^{\rm c}$Hardy\inst{4} \and P. Uttley\inst{5} }
\offprints{I. E. Papadakis;  e-mail: jhep@physics.uoc.gr}
\institute{Physics Department, University of Crete, P.O. Box 2208,
   710 03 Heraklion, Crete, Greece 
\and Max--Planck--Institut f\"ur extraterrestrische Physik,
   Giessenbachstrasse, D-85740 Garching, Germany
\and  MSSL, University College London, Holmbury St. Mary, Dorking,
   Surrey RH5 6NT, United Kingdom
\and School of Physics and Astronomy, University of Southampton, 
Southampton SO17 1BJ, United Kingdom
\and Sterrenkundig Instituut, Universiteit van Amsterdam, Kruislaan 403, 1098
SJ, Amsterdam, Netherlands} 
 
\date{Received ?/ Accepted ?} 
\abstract
{We present the results from the spectral analysis of the time-average spectrum
of the  Narrow Line Seyfert 1 (NLS1) galaxy Ark\,564 from a $\sim$ 100\,ks 
\xmm\ observation.}
{Characterize accurately the shape of the  time-average, X-ray continuum
spectrum of the source and search for any emission and/or absorption features
in it.}
{We use the \xmm\ data to obtain the X-ray spectrum of the 
source and we fit various spectral models to it.}
{The time-average, $3-11$ keV spectrum is well fitted by a power-law of slope
2.43. We detect a weak (equivalent width $\sim 80$ eV) emission line at $\sim 6.7$ keV, which
implies emission from ionized iron. There is no compelling evidence for
significant broadening of the line.We also detect a   possible Doppler shifted
absorption line at 8.1 keV. At energies $\la 2$ keV, the spectrum is dominated
by a smooth soft excess component which can be well fitted either by a two
black body components (kT$\sim 0.15$ and $0.07$ keV) or by a black body plus a 
relativistically blurred photoionized disc reflection model. We detect a broad,
shallow flux deficit in the $0.65-0.85$ keV band, reminiscent of the iron
unresolved transition array (UTA) features. We do not detect neither a strong
absorption edge around 0.7 keV nor an emission line around 1 keV.}
{The soft excess emission is consistent with being reflected emission from a
geometrically flat disc, with solar abundances, illuminated by an isotropic
source. The weakness of the iron line emission can be explained by relativistic
blurring. The UTA feature implies the presence of  warm material with a column
density of $2-5\times 10^{20}$ cm$^{-2}$. If the absorption line at 8.1 keV 
corresponds to Fe{\footnotesize XXVI} K$\alpha$, it suggests the presence of a
highly ionized absorbing material with N$_{\rm H}>10^{23}$ cm$^{-2}$,
outflowing at a high velocity of $\sim 0.17$c.}

\titlerunning{XMM$-$Newton observation of \ark}
\authorrunning{Papadakis \etal}
\maketitle
   
%%%%%%%%%%%%%%%%%%%%%%%%%%%%%%%%%%%%%%%%%%%%%%%%%%%%%%%%%%%%%%
\section{Introduction}
%%%%%%%%%%%%%%%%%%%%%%%%%%%%%%%%%%%%%%%%%%%%%%%%%%%%%%%%%%%%%%
\smallskip

Narrow-line Seyfert~1 (NLS1) galaxies are a peculiar group of active galactic
nuclei (AGN) characterized by their distinct optical  emission line properties
(Osterbrock \& Pogge 1985). In hard X-ray studies NLS1 galaxies comprise less
than 10\% of the Seyfert galaxies, however, from the ROSAT All-Sky Survey it
became clear that about half of the AGN in soft X-ray selected samples are NLS1
galaxies (Grupe 1996, Hasinger 1997). Boller, Brandt, \& Fink (1996) and Wang,
Brinkmann \& Bergeron  (1996)  found from \rosat\ observations that the soft
X-ray spectra of NLS1 galaxies are systematically steeper than those of broad
line Seyfert~1 galaxies. The popular explanation of the differences in X-ray
properties between Narrow and Broad line Seyferts is that the former objects
have relatively low black hole masses and high accretion rates (e.g. Pounds,
Done \& Osborne 1995, Puchnarewicz \etal\ 2001). 

Ark~564 is the X-ray brightest NLS1 galaxy with a 2$-$10~keV flux of $\sim
2\times10^{-11}$ \flux (Turner \etal\ 2001) and shows large amplitude
variations on short time scales (Leighly 1999a). \ark\ was observed for a
period of $\sim 35$ days in June/July 2000 by \asca\ as part of a
multi-wavelength AGN Watch monitoring campaign (Turner \etal\ 2001). The timing
behavior of the source was studied by Edelson \etal\ (2002), who found no
evidence of lags between any of the energy bands studied and that the
fractional variability amplitude was almost independent of energy band. Using
the same \asca\ data,  Papadakis  \etal\  (2002) reported a ``-1 to -2" slope
break in the power spectrum at high frequencies ($\sim 2\times10^{-3}$ Hz). On
the other hand, Pounds \etal\ (2001) detected a ``zero to -1" low frequency PSD
slope break at $\sim 1/13$ days$^{-1}$, using long term  \xte\ monitoring
observations. When combined, these two results support the idea of a small
black hole mass, and hence  high accretion rate, in \ark. Finally, Gliozzi
\etal\ (2002) found no statistically significant indications of
non-stationarity in the \asca\ light curves. Furthermore, using nonlinear
techniques they were able to demonstrate that the source behaves differently in
the high and low flux states. 

From the \asca\ long-look data the 2$-$10 keV X-ray spectrum was found to be 
quite steep ($\Gamma = 2.45 - 2.72$). It also showed a strong Fe K$\alpha$ 
line with an equivalent width of EW $\sim$ 350 - 650 eV, depending on the
fitted model, which  seemed to originate in highly ionized gas (Turner \etal\
2001). The presence of a soft excess was established but its exact form could
not be constrained. From a 50 ksec observation of \ark\ with the  {\it Chandra}
HETGS Matsumoto, Leighly \& Marshall (2004) confirmed the steep power-law
($\Gamma = 2.54\pm0.06$) above 2 keV and claimed the  detection of an edge-like
absorption feature at 0.712 keV.  An emission like feature at $\sim$ 1 keV has
been reported from various low-resolution spectra from \rosat, \asca\ and {\it 
BeppoSax}  (Brandt \etal\ 1994, Turner, George \& Netzer 1999, Comastri \etal\
2001), but its origin  remains unclear. Comastri \etal\ (2001) claimed the
detection of  a narrow iron emission line at $\sim$ 6.8 keV and an absorption
edge at $\sim$ 9.5 keV in the source rest frame from {\it BeppoSax}
observations. Vaughan \etal\ (1999) detected an edge at 8.5 keV in combined
\asca\ and \xte\ data, which they attributed to reflection from a strongly
irradiated disc. Finally,  Vignali et al. (2004) presented results from the
analysis of two short  (12 and $\sim$ 6 ksec) XMM-Newton observations. They
also found a steep  hard power-law ($\Gamma \sim 2.52 - 2.56$). The soft excess
was fitted with a black body component with a temperature of $\sim$ 140 eV plus
an absorption edge at $\sim$ 740 eV, with an optical depth, $\tau$, of  $\sim
0.4$.  For both pointings the quality of the  best fits was only moderate
($\chi^2_{red} \sim 1.2$)  despite of the limited photon statistics.

In this paper we present the results  from the analysis of the time-average
spectrum of the source resulting from  a 100 ksec XMM-Newton observation of
\ark. After describing the observation details, in Sect.~3 we present a short
analysis of the temporal behaviour of the source. In Sect.~4 and 5 we describe
in detail the spectral analysis of the data and we discuss the X-ray spectral
properties of \ark, respectively. Our conclusions are presented in Sect.~6.

%%%%%%%%%%%%%%%%%%%%%%%%%%%%%%%%%%%%%%%%%%%%%%%%%%%%
\section{Observation and Data Analysis}
%%%%%%%%%%%%%%%%%%%%%%%%%%%%%%%%%%%%%%%%%%%%%%%%%%%%%

Ark~564 was observed with XMM-Newton from 2005 January 5, 19:47 to 2005 January
6, 23:16 for 101,774 sec (obsID:0206400101). The PN and the two MOS cameras
were operated in Small Window mode with a medium filter. The EPIC data were
reprocessed  with the {\footnotesize XMMSAS}  version 6.5 and for the spectral
analysis we used the most recent versions of the response matrices.

The background count rate was very low (in total less than 0.6\% of the source 
count rate), apart from two small, short flares at the beginning of the
observation. Data from these periods were disregarded from the spectral
analysis.

With an average  count rate of $\sim$ 30 cts~s$^{-1}$ photon pile-up is
negligible for the PN detector, as  was verified using the {\footnotesize
XMMSAS} task $epatplot$. Source counts were accumulated  from 27$\times$26 RAW
pixels  (1 RAW pixel $\sim$ 4.1\arcsec) around the  position of the source. 
Background data were extracted from a similar, source free, region on the chip.
In order to minimize the effects of any calibration uncertainties, we selected
only single events for the spectral analysis (PATTERN$=0$ and FLAG=0; for
details of the instruments see Ehle et al. 2005) in the energy range from 300
eV to 12 keV. In total $\sim$ 2$\times10^6$ photons were accumulated in an
integration time of $\sim$ 69300 sec.

For the MOS data, with average count rates of $\sim$ 8 cts~s$^{-1}$, pile-up is 
not negligible. We therefore accumulated the source counts from  a ring  of
outer radius 45\arcsec~ excluding the innermost  12\farcs5 centered on the 
position of \ark. The background data were extracted from a similar region
from another chip. Events with  PATTERN $\leq$ 12 and FLAG=0 were used for the
analysis. In a net exposure time of $\sim$ 96 ksec, $\sim 2.3\times10^5$ 
photons were selected for each of the two MOS cameras in the 0.3$-$10 keV
energy band. 

Finally, the RGS data were reduced using {\footnotesize RGSPROC} in
{\footnotesize XMMSAS} 6.5 and the latest available calibration files (November
2005). Residuals in the current effective area calibration were corrected using
the data/model ratio of a power-law fit to the rev 0084 observation of the
continuum source Mrk~421. The first and second order spectra and response
matrices from RGS1 and RGS2 were resampled to the first order RGS1 spectrum,
then combined to produce a single spectrum and a single response matrix.

%%----------------------------- FIGURE 1----------------------------
\begin{figure}
\psfig{figure=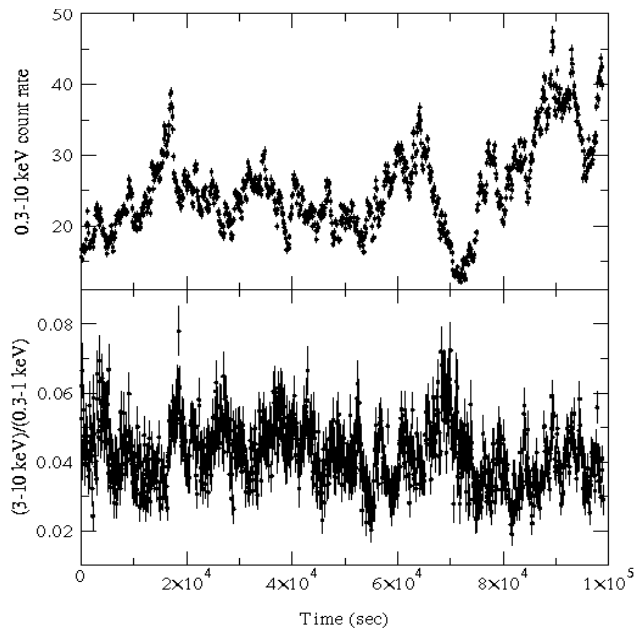,height=8.3truecm,width=8.3truecm,angle=0,%
 bbllx=170pt,bblly=278pt,bburx=425pt,bbury=562pt,clip=}
\caption[]{Upper panel: background subtracted PN light curve of  Ark 564 in
the  0.3$-$10 keV energy band. Note that in reality the average count rate is
slightly larger, as we have not  corrected the count rate for the 71\%  live
time of the Small Window mode of the detector. Lower panel: the corresponding
(3$-$10\,keV)/(0.3$-$1\,keV) hardness ratio curve. The time binning in both
panels is 100 sec.}
\label{figure:lc}
\end{figure}

%%-------------------------------- FIGURE 2 ----------------------------
\begin{figure}
\psfig{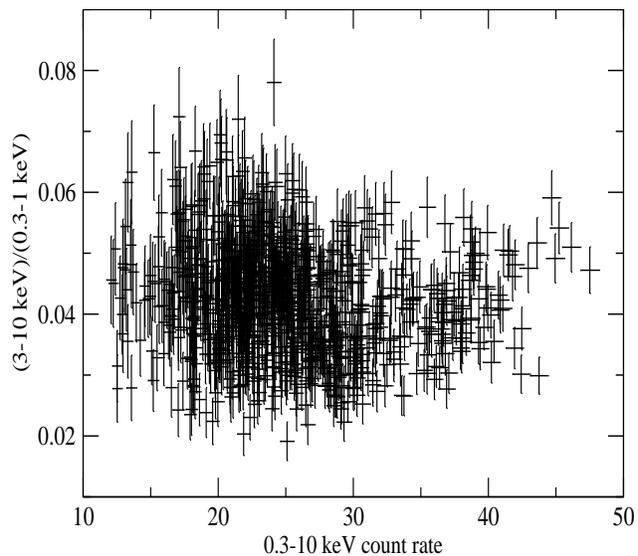}
 \caption[]{The (3$-$10\,keV)/(0.3$-$1\,keV) hardness ratio plotted as a
function of the total count rate.}
\label{figure:hrcts}
\end{figure}

%%%%%%%%%%%%%%%%%%%%%%%%%%%%%%%%%%%%%%%%%%%%%%%%%
\section{Timing analysis}
%%%%%%%%%%%%%%%%%%%%%%%%%%%%%%%%%%%%%%%%%%%%%%%%%%%%%%%%

Ark~564 is a highly variable source. A detailed timing analysis of the observed
light curves in various energy bands will be presented in  M$^{\rm c}$Hardy
\etal\ (in preparation). Recently, Arevalo \etal\ (2006) combined the data from
the 100 ks  \xmm\ observation with those from the  month-long monitoring \asca\
observation to calculate the time lags and coherence functions between various
energy bands. 

The upper panel of  Fig.~\ref{figure:lc} shows the  PN 0.3$-$10 keV background
subtracted light curve, binned in 100 sec intervals. The source is highly
variable on all sampled time scales. The max/min variability amplitude is of
the order of $\sim 4$, while the fractional variability amplitude (corrected
for the experimental noise) is $f_{\it rms} = 24.8\pm 0.1$\% (the error
accounts only for the  measurement error in the light curve points, and has
been estimated according to the prescription of Vaughan \etal\ (2003)).  We
also considered light curves in 3$-$10\,keV and 0.3$-$1\,keV as
representative of the ``hard" and ``soft" energy bands, respectively. The hard
band variations are of larger amplitude ($f_{\rm rms,hard}=27.2\pm 0.4$\% as
opposed to $f_{\rm rms,soft} = 24.7\pm 0.1$\%). This difference suggests the
presence of spectral variations. 

In the lower panel of Fig.~\ref{figure:lc} we plot the
(3$-$10\,keV)/(0.3$-$1\,keV) hardness ratio as a function of time.  This plot 
reveals that the observed flux variations are indeed associated with spectral
variations as well. Although significant, they are of rather small amplitude. 
In some cases, like for example around $\sim 18$ and 75 ksec from the start of
the observation, the spectrum becomes clearly harder as the flux decreases.
However, it is not obvious that this trend holds at all times. In fact, the
``hardness ratio versus count rate" plot (Fig.~\ref{figure:hrcts}) demonstrates
clearly that the  relation between flux and spectral variations  is far from
clear and simple. The results from a detailed analysis of the observed spectral
variations will be presented in a subsequent paper (Brinkmann \etal, in
preparation). 

Using the results from the spectral analysis presented below, the average
0.3$-$10 keV flux amounts to $\sim 1\times10^{-10}$ erg\,cm$^{-2}\,$s$^{-1}$ 
which results in an X-ray luminosity of L$_{0.3-10\,keV} = 6.5\times 10^{43}$
erg\,s$^{-1}$,  assuming a redshift of $z=0.0247$ and a Friedman cosmology with
H$_o$ = 75 km\,s$^{-1}$Mpc$^{-1}$, q$_o$=0.5. A substantial part of the flux is
emitted in the soft energy band. The average 2$-$10 keV flux  amounts to
1.5$\times10^{-11}$ erg\,cm$^{-2}\,$s$^{-1}$ which is comparable to  the mean
flux of 2$\times10^{-11}$ erg\,cm$^{-2}\,s^{-1}$ during the 35 day long \asca\
observation in June 2000 (Turner \etal\ 2001).

%%%%%%%%%%%%%%%%%%%%%%%%%%%%%%%%%%%%%%%%%%%%%%%%%%%%%%%%%%%%%%%%%
\section{Spectral analysis of the PN and MOS data}
%%%%%%%%%%%%%%%%%%%%%%%%%%%%%%%%%%%%%%%%%%%%%%%%%%%%%%%%%%%%%%%%%

For the spectral analysis the source counts were grouped with a minimum of 30
counts per energy bin. Spectral fits have been performed with the
{\footnotesize XSPEC v11.3} package. Spectral responses and the effective area
for the PN and MOS spectra were generated with the {\footnotesize SAS} commands
{\em rmfgen} and {\em arfgen}. The errors on the best-fitting model parameters
represent the 90\% confidence limits for one interesting parameter, and in the
cases we report upper limits, these correspond to the 99\% confidence limit.
The energy of the emission or absorption features are given in the rest frame
of the source. Finally,  we consider a model as providing an acceptable fit to
the data if the goodness of fit is at better than the 5\% confidence level, and
we accept that the addition of a  model component is necessary if the quality
of the model fitting is improved at more than the 95\% significance level. 

%%=================================================
\subsection{The hard band EPIC PN spectrum}
%%=================================================
 
The spectral complexity of \ark\ in the $0.3-10$ keV band is well known from
previous observations. We therefore started by fitting the hard band, i.e.
3$-$11 keV, PN spectrum  with a simple power-law. The Galactic absorption was
modeled using {\footnotesize PHABS} in {\footnotesize XSPEC} and the abundance
table of Lodders (2003), keeping the value of the interstellar absorption fixed
at  N$_{\rm H}$ =  6.4$\times10^{20}$ cm$^{-2}$ (Dickey \& Lockman 1990). 

\begin{figure}
\psfig{figure=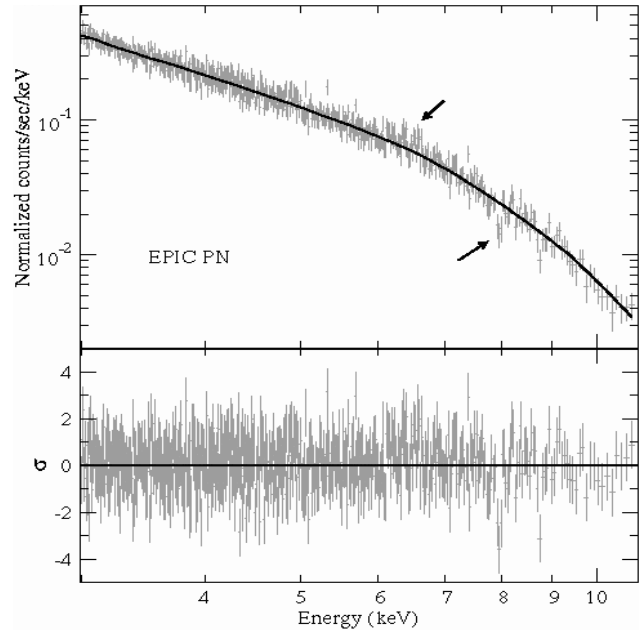,height=8.3truecm,width=8.3truecm,angle=0,%
 bbllx=172pt,bblly=292pt,bburx=413pt,bbury=564pt,clip=}
\caption[]{Plot of the power-law fit (upper panel) and residuals (lower panel) 
to the PN data in the $3-11$ keV energy band. Notice the residuals in the Fe
line region and  the absorption feature around $\sim 8$ keV (indicated with the
arrows in the upper panel). In this, and all subsequent similar plots, the best
model fitting residuals are plotted in terms of $\sigma$ with error bars of
size one.}
\label{figure:pn-hard}
\end{figure}

The fit is acceptable with a reduced $\chi^2_{\rm red}$/dof = 0.987/631. The
best fitting results are listed in Table~1 (model {\footnotesize ``PL"}).
Fig.~\ref{figure:pn-hard} shows the best {\footnotesize PL} model fit to the PN
data  and the corresponding  residuals. The only apparent systematic
deviations  (indicated with arrows in Fig.~\ref{figure:pn-hard}) appear in
$\sim 6-6.5$ keV (where iron line emission features are expected) and around 8
keV, where a narrow absorption feature appears. Even if real, it is obvious
that these features are rather weak. 

%%----------------------------- figure 3 -----------------------------

We repeated the fit with the addition of a Gaussian emission line.  Initially
we kept the width of the line fixed at $\sigma=100$ eV (i.e. smaller than the
PN's resolution at these energies). The results are listed in Table~1 (model
{\footnotesize ``PL+NGL"}). The addition of the narrow line improves the
goodness of fit by $\Delta \chi^2=13$ for 2 dof,  significant at the $99.9$\%
level. The line's energy of  $6.67^{+0.08}_{-0.10}$~keV is indicative of
emission from Fe\,{\footnotesize XXV}. Inspection of the residuals suggests  a
possible second emission line feature at $\sim 6.4$ keV. Consequently, we added
a second Gaussian line, keeping the energy of the first line fixed at its best
fitting value of 6.67 keV. We list the results in Table~1 (model {\footnotesize
``PL+2\,NGLs"}). The quality of the fit improves only  by $\Delta \chi^2=3.4$
for 1 dof, which is not significant.   We then examined whether the fit would
improve if we let the width of the 6.7 keV emission line vary as a free
parameter.  This improves the fit by $\Delta \chi^2=2$ for 1 dof, which again
is not significant. The best fitting $\sigma$ value of $\sim 0.2$ keV (Table~1,
model {\footnotesize ``PL+BGL"}) is only slightly larger than the PN's
intrinsic resolution of $\sim 150-170$ eV at $6-7$ keV. 

We then repeated the model fitting with a {\footnotesize DISKLINE} model
(Fabian \etal\ 1989) substituted for the Gaussian emission line.  Since the
emission feature in the PN spectrum is not strong, we kept the inner and outer
disc radii fixed to 10 and 1000 gravitational radii ($r_{g}$), respectively,
the emissivity index to $-2$, and the inclination angle to 30 degrees. The best
fitting results are listed in Table~1 (model ``{\footnotesize PL+DL}"). The
goodness of fit is similar to that of the  {\footnotesize PL+NGL} model, and
the best best-fitting line energies are similar in the two models. 

\begin{figure}
\psfig{figure=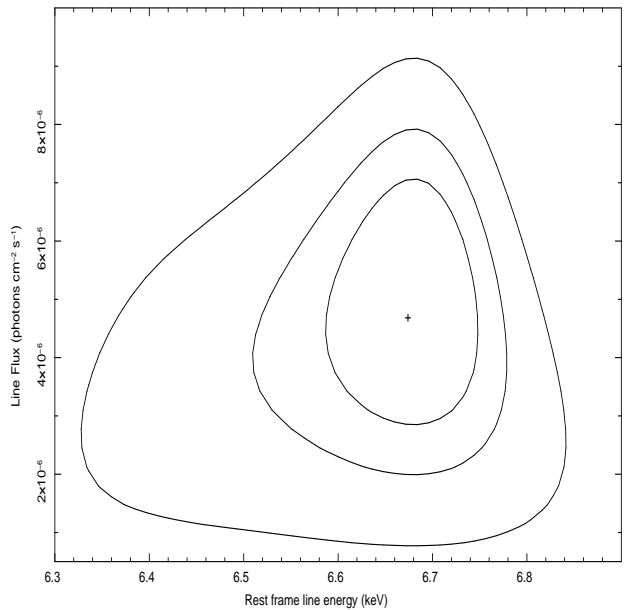,height=8.3truecm,width=8.3truecm,angle=-90,%
bbllx=55pt,bblly=37pt,bburx=563pt,bbury=706pt}
\caption[]{Plot of the 68\%, 90\%, and 99\% confidence level contours (from the
inner to the outer curves, respectively) for the rest-frame energy
and the line flux of the iron emission line in the case of the {\footnotesize
PL+NGL} best fitting model to the EPIC PN data.}
\label{figure:contour}
\end{figure}

We conclude that there is no significant evidence for the presence of a broad
iron emission line in the  EPIC PN data. In Fig.~\ref{figure:contour}  we show
the confidence contour plot for the rest-frame energy and intensity of the
detected line in the case of the {\footnotesize PL+NGL} model. The contours
plotted correspond to the 68\%, 90\%, and 99\% confidence level (from the inner
to the outer curves) for two interesting parameters.

%%----------------------------- figure 4 -----------------------------

\begin{figure}
\psfig{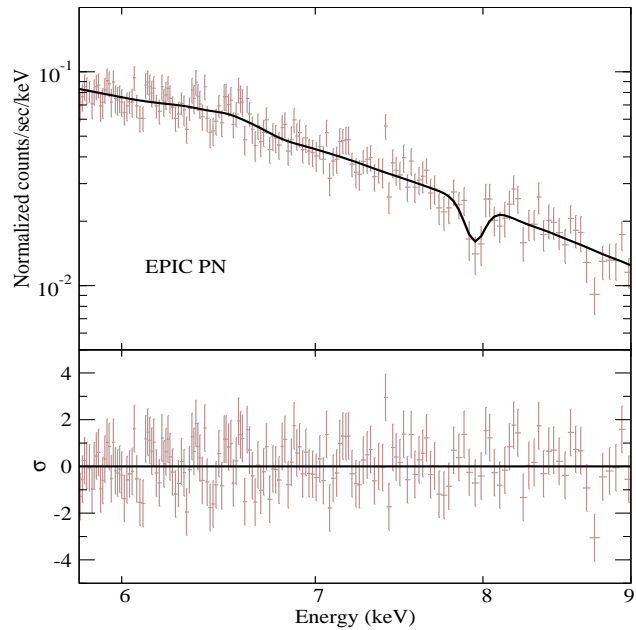}
\caption[]{``Power-law plus discline plus absorption line"  best fitting model
plot to the PN data (upper panel) and residuals plot (lower panel) 
in the $5.8-9$ keV energy band. }
\label{figure:pn-hard2}
\end{figure}

In order to assess the importance of the absorption feature at 8 keV,  we added
a narrow Gaussian line (i.e. with $\sigma$ kept fixed at 10 eV)  in the
{\footnotesize PL+DL} model and forced its normalization to have negative
values only (model ``{\footnotesize PL+DL+ABL}" in Table~1). Compared to the
best fitting {\footnotesize PL+DL} model, this addition  improved the goodness
of fit by $\Delta \chi^2=17.7$  for 2 dof, which is highly significant. The
best-fitting absorption line's energy is  $\sim 8.1$ keV, and its equivalent
width is $\sim -60$ eV.  Fig.~\ref{figure:pn-hard2} shows the best fitting
{\footnotesize PL+DL+ABL}  plot to the  PN. The residuals plot (lower panel in
the same figure) shows no clear systematic deviations from this model. The
model residuals in the region between $6-6.5$ keV are somehow ``noisy", but
they are almost certainly caused by local statistical fluctuations. Further
attempts to model them with extra narrow emission or absorption components did
not lead to any reasonable results.

%%--------------------------------- TABLE 1 ----------------------------
\begin{table*}
\tabcolsep1ex
\caption{\label{hardfit} Results from spectral fitting in the $3-11$ keV energy
band assuming fixed Galactic N$_{\mathrm H} = 6.4\times10^{20}$cm$^{-2}$.
\hfill \break The
letter $f$ indicates  parameters whose value was kept fixed during the model
fitting.}
\begin{tabular}{llcccccccc}
\noalign{\smallskip} \hline \noalign{\smallskip}
 & Model & $\Gamma$ & E$_{\rm{line}}(keV)$ & $\sigma$(keV) & EW(eV) & 
 E$_{\rm{line,2}}(keV)$ & $\sigma_2$(keV) & EW$_2$(eV) &
$\chi^{2}_{\rm{red}}$/dof \\
\noalign{\smallskip} \hline \noalign{\smallskip}

PN & PL & $2.44\pm0.03$ & $-$ & $-$ & $-$ & $-$ & $-$ & $-$ & 0.987/631 \\

PN & PL+NGL & $2.46\pm 0.03$ & $6.67^{+0.08}_{-0.10}$ 
&$f(0.1)$ & $44\pm 20$ & $-$ & $-$ & $-$ & 0.97/629 \\

PN & PL+2\,NGLs & $2.47\pm 0.03$ & $f(6.67)$ & $f(0.1)$ & $40\pm 20$ & 
 $6.35^{+0.13}_{-0.16}$ & $f(0.1)$ & $23^{+17}_{-19}$ & 0.966/628 \\

PN & PL+BGL & $2.47\pm 0.03$ & $6.60^{+0.14}_{-0.15}$ 
&$0.22^{+0.38}_{-0.17}$ & $66^{+60}_{-32}$ & $-$ & $-$ & $-$ & 0.968/628 \\

PN & PL+DL & $2.47\pm 0.03$ & $6.59^{+0.08}_{-0.09}$ & $-$ 
& $76\pm 30$ & $-$ & $-$ & $-$ & 0.965/629 \\

PN & PL+DL+ABL & $2.45\pm 0.03$ & $6.6\pm0.1$ & $-$ 
& $67\pm 30$ & $8.14\pm 0.04$ & $f(0.01)$ & $-54\pm 20$ & 0.94/627 \\

MOS1+2 & PL & $2.34\pm0.05$ & $-$ & $-$ & $-$ & $-$ & $-$ & $-$ & 1.199/257 \\

MOS1+2 & PL+NGL & $2.36\pm 0.05$ & $7.04^{+0.09}_{-0.11}$ 
&$f(0.1)$ & $81^{+49}_{-43}$ & $-$ & $-$ & $-$ & 1.171/255 \\

MOS1+2 & PL+2\,NGLs & $2.38\pm 0.05$ & $f(7.04)$ & $f(0.1)$ & $83\pm 46$ & 
 $6.64^{+0.22}_{-0.28}$ & $f(0.1)$ & $38\pm 35$ & 1.163/254 \\

MOS1+2 & PL+BGL & $2.38\pm 0.05$ & $6.89\pm 0.26$ 
&$0.29^{+0.26}_{-0.11}$ & $138^{+54}_{-70}$ & $-$ & $-$ & $-$ & 1.172/254 \\

MOS1+2 & PL+DL & $2.39\pm 0.05$ & $6.76\pm 0.10$ & $-$  & $158^{+71}_{-77}$ &
$-$ & $-$ & $-$ & 1.159/255 \\

MOS1+2 & PL+DL+ABL & $2.37\pm 0.05$ & $6.76\pm0.11$ & $-$  & $141^{+75}_{-72}$
& $7.91\pm 0.08$ & $f(0.01)$ & $-62\pm 42$  & 1.146/253 \\

\noalign{\smallskip}\hline
\end{tabular}
\end{table*}

%%------------------------------ figure 5 ----------------------------------

%%============================================================
\subsection{The hard band EPIC MOS energy spectra}
%%=============================================================

\begin{figure}
\psfig{figure=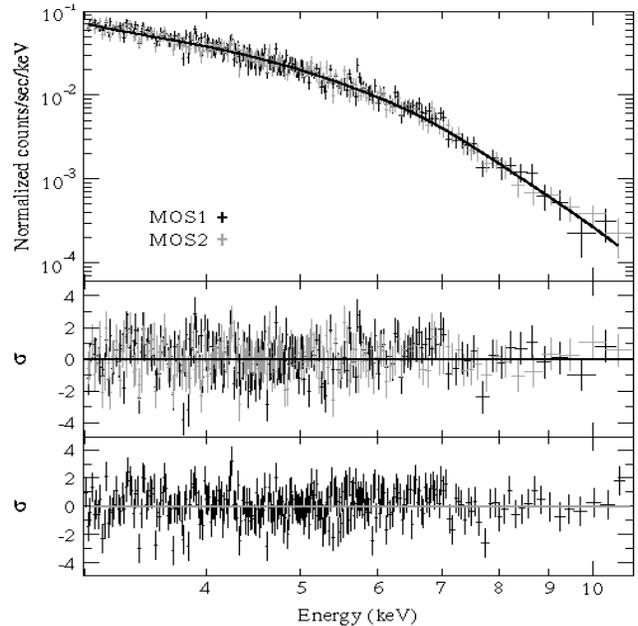,height=8.3truecm,width=8.3truecm,angle=0,%
 bbllx=172pt,bblly=277pt,bburx=417pt,bbury=562pt,clip=}
\caption[]{Plot of the power-law fit (upper panel) and residuals (middle
panel)  to the joint MOS 1 and 2 data in the $3-11$ keV energy band. In the
bottom panel we plot the best power-law fit residuals in the case of the
combined MOS1+2 spectrum.  Notice the residuals in the Fe line region and the
absorption feature around $\sim 7.8$ keV, just like in the case of the PN
spectrum.}
\label{figure:mos-hard}
\end{figure}

Fig.~\ref{figure:mos-hard} shows the best-fitting PL model fit to the MOS1 and
2 data and the corresponding residuals in the $3-11$ keV band (top and middle
panels, respectively). Clearly, the MOS 1 and MOS 2 spectra in the hard band
are similar. For that reason, and in order to increase the signal to noise
ratio, we added them together and created the combined ``MOS1+2" spectrum. The
bottom panel in the same Figure shows the residuals in the case of the
best-fitting PL model to the MOS1+2 spectrum. The residuals plot is rather
noisy but a {\footnotesize PL} model appears to fit the overall spectrum well.
The main residual feature is  a low-amplitude excess emission around $\sim
6.5-7$ keV.

We fitted the MOS1+2 spectrum with the same models that we used in the case of
the PN spectrum. The best fitting results are listed in Table~1  and are
broadly  consistent with those we reached from the PN spectral analysis.
Compared with the best fitting {\footnotesize PL} model,  the addition of
either a narrow, or  a broad Gaussian line,  or a {\footnotesize DISKLINE}
component improves the quality of the fit at the $98.1, 96.6$ and $99.5$\%
levels, respectively. Compared to the best fitting {\footnotesize PL+NGL}
model, neither the addition  of a second narrow Gaussian line nor the use of
the {\footnotesize PL+BGL} model improves the fit significantly.

There is a hint of an absorption line at 7.9 keV in the MOS1+2 spectrum,  but
the  statistics of the MOS data are not sufficient to confirm or refute the
absorption line that appears in the PN data at 8.1 keV.  Indeed, the
improvement to the goodness of fit when we add an absorption line to the
{\footnotesize PL+DL} model is not significant (it improves the fit by 91.3\%).
This is not surprising given the poorer statistics of the MOS spectra with
respect to the PN spectrum. However, note that, among all the models listed in
Table 1, only the  {\footnotesize PL+DL+ADL} model fits the MOS1+2 spectrum  at
better than the 5\% confidence level. 

As commonly found, the fitted  power-law slopes for the MOS are slightly
flatter than those for the PN by $\Delta\Gamma \sim 0.08-0.1$, for all  models
listed in Table~1. Furthermore, the MOS1+2  best fitting line centroid energies
are systematically higher, and the line's EW larger, than those resulting from
the PN model fitting. If we consider  the {\footnotesize PL+DL+ABL} model, for
which the  differences in the PN and MOS1+2  best model fitting parameter
values are the smallest,  then the average PN+MOS spectral slope is
$\Gamma=2.43\pm0.03$,  the average iron emission line energy is $6.67\pm 0.07$
keV, and its equivalent width is $78\pm28$ eV.

We conclude that both the PN and MOS $3-11$ keV band  spectra are well fitted
by a power-law model of slope $\Gamma\sim 2.43$. There is significant evidence
for an iron emission line at $6.7$ keV which implies emission from ionized
iron. We find no evidence for an extra narrow line emission component either at
lower or higher energies. The line is weak (EW$\sim 80$ eV) and narrow.
Although models like a broad Gaussian and a {\footnotesize DISKLINE} do fit the
data well, they are not statistically required. Finally, a narrow absorption
feature at $8.1$ keV appears in the PN spectrum. 

%%----------------------------- figure 6 -----------------------------

%%------------------------------ figure 7 ----------------------------------
\begin{figure}
\psfig{figure=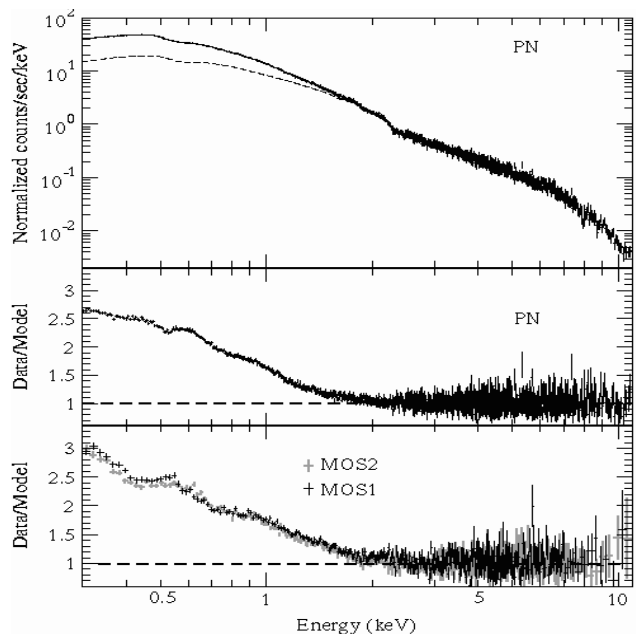,height=8.3truecm,width=8.3truecm,angle=0,%
 bbllx=172pt,bblly=270pt,bburx=417pt,bbury=564pt,clip=}
\caption[]{Extrapolation of the hard band, best fitting ``Pl+DL+ABL" model into
the soft band to demonstrate the soft excess. The upper two panels show the
full-band data, the hard band best model fit and the data-to-model ratio  for
the PN data, while the lower panel shows  the ratios for the two MOS spectra.} 
\label{figure:extrapol}
\end{figure}

%%====================================================
\subsection{The broad band EPIC spectra}
%%===================================================

An extrapolation of the best fitting {\footnotesize PL+DL+ABL}  model to low
energies shows a broad excess  of emission, both in the PN and the two MOS
spectra (see Fig.\,\ref{figure:extrapol}).  The lower panels in
Fig.\,\ref{figure:extrapol} show the data/model residuals for the PN and MOS
spectra. The smooth, extra component which dominates  the source's emission at
energies below 2 keV is clearly seen. There appear no strong spectral features,
either in emission or absorption, while the soft excess component flattens   at
energies below $\sim 0.5$ keV. 

The nature of the small amplitude structures that we observe might be intrinsic
to the source but remaining calibration uncertainties cannot be ruled out
either, considering the excellent signal to noise that we have achieved at low
energies in this long observation. For example, the obvious structures at
$\sim$ 0.5\,keV are most probably caused by calibration uncertainties. 
Furthermore, the MOS 1 and 2 spectra, although quite similar at energies above
$\sim 0.6$ keV, deviate significantly at lower energies. Clearly, this also
reflects remaining calibration uncertainties in the MOS detectors at low
energies.

In order to minimize the influence of the instrumental response and  gain some
insight into the intrinsic  broad-band shape of the source spectrum, the PN
data of \ark\ were compared to the PN spectrum of 3C~273. Specifically, we
divided the PN raw spectrum by the raw spectrum of a  $\sim$ 18 ksec
observation of 3C\,273 from June 30, 2004, taken in the same SW mode with
medium filter. We chose 3C~273 as it is a bright source with a relatively
simple spectrum in the EPIC band, i.e. it shows a hard power-law plus smooth
soft excess modified by Galactic absorption,  without strong, sharp spectral
features such as edges (Page \etal\ 2004). As a result, the ratio of the \ark\
to the  3C~273 spectrum will factor out most effects of the instrumental
response and will give a better view of the intrinsic shape of the \ark\
spectrum.

\begin{figure}
\psfig{figure=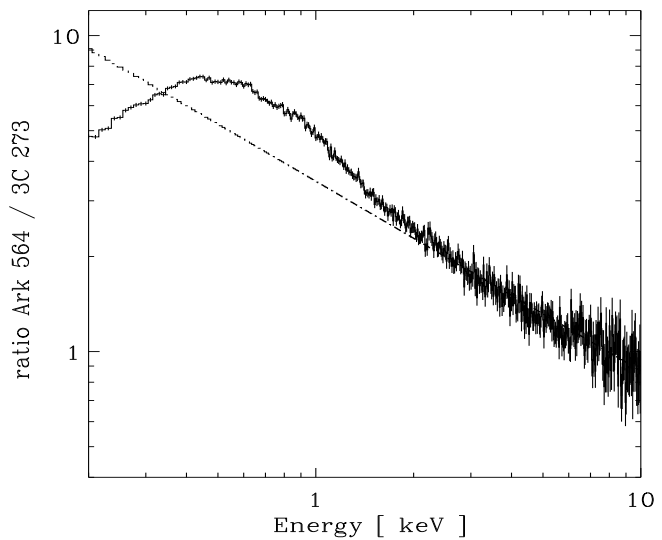,height=7.3truecm,width=8.7truecm,angle=0,%
 bbllx=85pt,bblly=362pt,bburx=424pt,bbury=702pt,clip=}
\caption[]{Ratio of the raw PN counts of \ark\ divided by the raw PN counts of
3C 273. The dash-dotted line is a power-law approximation to the hard part of
the spectrum with a slope of $\Gamma = 0.6$ representing the difference of the
slopes of the hard spectra of the two sources.}
\label{figure:ratio}
\end{figure}

In Fig.\,\ref{figure:ratio} we display this ratio. The straight dash-dotted
line indicates the difference in the slopes of the intrinsic hard band  power
laws of the two objects. The downward bending of the ratio at energies below
$\sim 0.4$ keV  can be attributed to the smaller amount of absorption for
3C\,273 (N$_{\rm H} \sim 1.69\times10^{20}$cm$^{-2}$). The ratio shows some
small amplitude  ``wiggles'' but by no means any prominent spectral structures.
For example,  Vaughan \& Fabian (2004) have followed a similar approach, and
show in their Fig.~3 the ratio of the raw MCG -6-30-15 PN spectrum to the raw
PN spectrum of 3C~273. We do not observe any of the strong absorption features
that appear in MCG -6-30-15 at low energies. In fact, the ratio plot shown in
Fig.\,\ref{figure:ratio} suggests that both  the existence of  strong
absorption edges and/or emission lines in the $\sim 0.5-1.5$ keV band  is
rather unlikely. Furthermore, the ratio behaves smoothly even at the lowest
energies (down to 0.2\,keV), in contrast to the results from the spectral fits
(to the MOS spectra mainly) that we present in the following sections. 

%%------------------------------ figure 8 ----------------------------------

%%====================================================
\subsection{Describing the soft excess}
%%====================================================

We tried to model the soft excess with a variety of different models to
understand its physical nature. To this end, we made the following assumptions.
As with the hard band model fits,  the Galactic absorption was modeled using
{\footnotesize PHABS} with  N$_{\rm H}$ =  6.4$\times10^{20}$ cm$^{-2}$ and the
abundance table of Lodders (2003). We also tried the tables of  Grevesse \&
Sauval (1998) and  Wilms, Allen \& McCray (2000), but their use resulted in
systematically worse best model fits (for the adopted N$_{\rm H}$ value). 

For the PN spectrum, we used the best fitting {\footnotesize PL+DL+ABL} model
as the ``baseline" model to which various components were added in order to
model the soft excess. The reason for this choice is not based on physical
considerations (i.e. the iron emission feature in the hard band PN  spectrum is
equally well fitted by a narrow Gaussian line) but simply on the fact that this
model yields  the smallest best fitting $\chi^{2}$ value, in the hard band,
among all the models listed in Table~1. During the model fitting, the model
parameter values  were kept fixed to the values listed in Table 1, except for
the {\footnotesize PL} normalization which was left as a free parameter. 

\begin{figure}
\psfig{figure=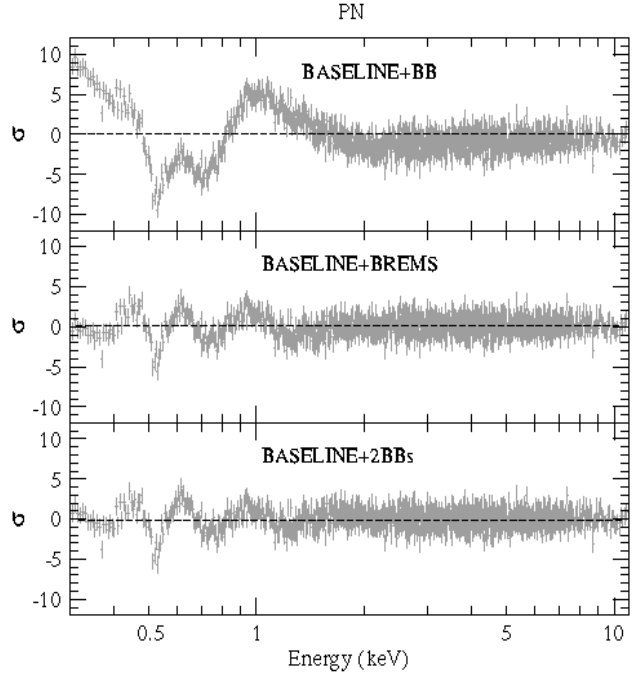,height=9.0truecm,width=8.3truecm,angle=0,%
  bbllx=169pt,bblly=322pt,bburx=410pt,bbury=564pt,clip=}
\caption[]{Best model fit residuals to the full band PN energy spectrum of Ark
564.}
\label{figure:softpn}
\end{figure}

The models we employed for the soft band spectrum were the following: 1) a soft
band power-law (``{\footnotesize BASELINE+PL}"), 2) a break in the hard power 
law (``{\footnotesize BKN PL}"), 3) two breaks in the hard power-law
(``{\footnotesize 2\,BKN PL}"),  4)  a blackbody (``{\footnotesize
BASELINE+BB}"), 5) two blackbodies (``{\footnotesize BASELINE+2\,BBs}" ),  and 
6) a thermal bremsstrahlung spectrum (``{\footnotesize BASELINE+BREMS}"). None
of them resulted in  statistically acceptable fits. In Table~2 we list the
$\chi^{2}_{\rm{red}}$/dof for each  model, and in Fig.\,\ref{figure:softpn} we
show the best fitting model residuals in the case of the most favorable model
fits to the PN spectrum ({\footnotesize BASELINE+2\,BBs} and {\footnotesize
BASELINE+BREMS}). For comparison reasons, we also show the best fitting
residuals in the case of the {\footnotesize BASELINE+BB} model.

Despite the formally unacceptable large $\chi^{2}_{red}$ values, the plots in
Fig.\,\ref{figure:softpn} suggest that the {\footnotesize BASELINE+2\,BBs} and
{\footnotesize BASELINE+BREMS} models provide reasonable fits to the broad band
PN spectrum. In the case of the {\footnotesize BASELINE+BREMS} model  we find
kT$\sim 0.3$ keV. The best-fitting temperatures of the two blackbodies are
$\sim 0.08$ and $\sim 0.16$ keV  (since the models do not give acceptable fits,
error estimates of the model parameters are not meaningful). 

%%--------------------------------- TABLE 2 ----------------------------
\begin{table}
\caption{\label{hardfit} Results from the  spectral fitting in the $0.3-11$ keV
energy band assuming fixed Galactic N$_{\mathrm H} =
6.4\times10^{20}$cm$^{-2}$.}
\begin{tabular}{lcc}
\noalign{\smallskip} \hline \noalign{\smallskip}
Model & $\chi^{2}_{\rm{red}}$/dof (PN) & $\chi^{2}_{\rm{red}}$/dof (MOS)\\
\noalign{\smallskip} \hline \noalign{\smallskip}
BASELINE+PL  & 11.85/1174 & 2.66/753\\
BKN PL & 4.84/1173 & 1.98/758 \\
2\,BKN PL & 1.52/1171 & 1.66/756 \\
BASELINE+BB  & 4.88/1173 & 4.76/753 \\
BASELINE+2\,BBs &  1.29/1171  & 1.55/751 \\
BASELINE+BREMS & 1.35/1173 & 1.81/753 \\
\noalign{\smallskip}\hline
\end{tabular}
\end{table}

%------------------------------ figure 9 ----------------------------------

%%------------------------------ figure 10 ----------------------------------
\begin{figure}
\psfig{figure=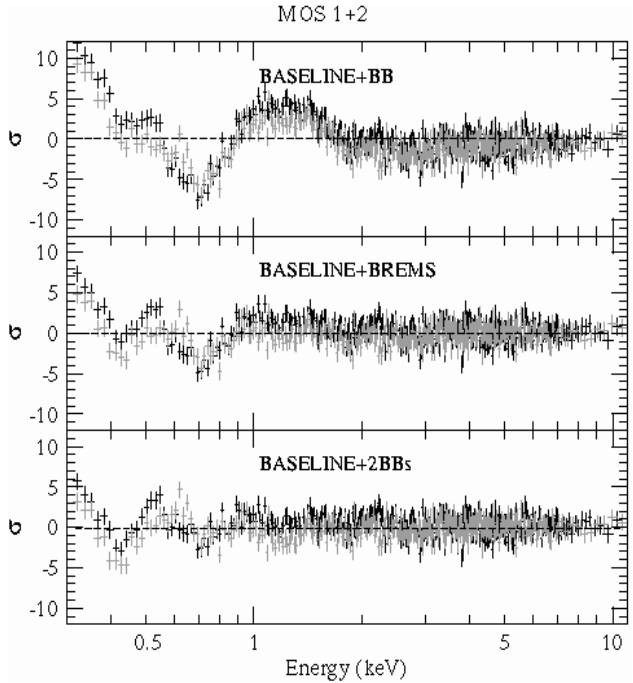,height=9.0truecm,width=8.3truecm,angle=0,%
  bbllx=170pt,bblly=324pt,bburx=410pt,bbury=564pt,clip=}
\caption[]{Best model fit residuals to the full band MOS 1 and 2 energy spectra of
Ark 564.}
\label{figure:softmos}
\end{figure}

The models listed in Table~2 were also fitted to the  MOS 1 and 2 spectra.  In
this case, we did not add the two MOS spectra together, due to their
differences at low energies. Instead, we performed joint model fits to both of
them, and we plot the best model fits residuals for both detectors.  The best
fitting $\chi^{2}_{red}$ results are listed in the last column of Table 2 and
residual plots for a few models  are plotted in Fig.\,\ref{figure:softmos}. As
with the PN spectrum, none of the models provide a statistically acceptable fit
to the data. The {\footnotesize BASELINE+2\,BBs} and {\footnotesize
BASELINE+BREMS} models  provide again reasonable fits to the MOS spectra. The
best fitting parameter values (kT$\sim 0.37$ keV, in the case of the
{\footnotesize BASELINE+BREMS} model, and blackbody temperatures of $\sim 0.08$
and $0.19$ keV) are  consistent with the results from the PN spectrum.

A comparison between the respective plots in Fig.\,\ref{figure:softpn} and  
Fig.\,\ref{figure:softmos} shows that the residual structures are similar in
all detectors. The use of a single black body component always gives rise to a
broad excess bump in the $0.8-1.5$ keV range and deep, edge-like structures,
between $\sim 0.5$ and $0.8$ keV. Adopting a thermal bremsstrahlung or two
black body components considerably improves the fit, and minimizes the
residuals around the 1\,keV and $0.5-0.8$ keV features. At energies below 0.6
keV, the MOS1 and 2 residuals are different from each other and  from those
seen in the PN spectra.  These discrepancies must be attributed to calibration
uncertainties of the instruments. 

Clearly, as the reduced $\chi^{2}$ values demonstrate, an ideal bremsstrahlung
or two black body components are not exact models for the soft excess. However,
the deviations of the observed spectrum from the best fitting  model shapes are
only at a few percent level (see top panel in Fig.\,\ref{figure:finalratio})
and they could reflect to a large extent remaining calibration uncertainties. 
For example, the most prominent residual features in the PN best fitting 
{\footnotesize BASELINE+2\,BBs} and {\footnotesize BASELINE+BREMS} models are
around $\sim 0.5$ keV (where we expect the instrumental and Galactic oxygen
edges to appear). However, the other residual features around $0.7-0.8$ keV, 
where a broad, shallow deficit appears, and  $\sim 0.9-1$ keV, where we observe
an excess above the best fitting model could  correspond to low amplitude,
absorption features intrinsic to the source. This possibility can be 
investigated with the study of RGS spectrum, which we describe below. 

%========================================================================
\subsubsection{The RGS data}
%==========================================================================
 
Due to the EPIC-RGS cross calibration problems, and the time-dependent
degradation of the RGS low-energy effective area, which is not yet accounted
for in the {\footnotesize SAS}  (Stuhlinger \etal\ 2006), the RGS data cannot be
compared directly to the best fitting EPIC models. However, this is not a
serious problem, as our main aim in this work is to identify any strong
emission and/or absorption features in the RGS spectra that may affect the
goodness of the various model fits to the EPIC spectra.  To this end, all we
need is to model the continuum in the RGS spectrum as accurately as possible
and search for any remaining residuals.

We therefore used the {\footnotesize BASELINE+2BBs} model to fit the RGS
spectrum. The power-law and blackbody model parameters were left to vary as
free parameters and the Galactic absorption was modeled as in the case of the
EPIC spectra. The best-fitting RGS  power-law index is harder  than that found
with PN ($\Delta \Gamma\sim-0.5$), as expected with the current  RGS/EPIC
cross-calibration  (Stuhlinger \etal\ 2006). The two blackbodies have best
fitting kT of $0.1$~keV and $0.2$~keV, similar to those  determined with the
PN.

\begin{figure}
\psfig{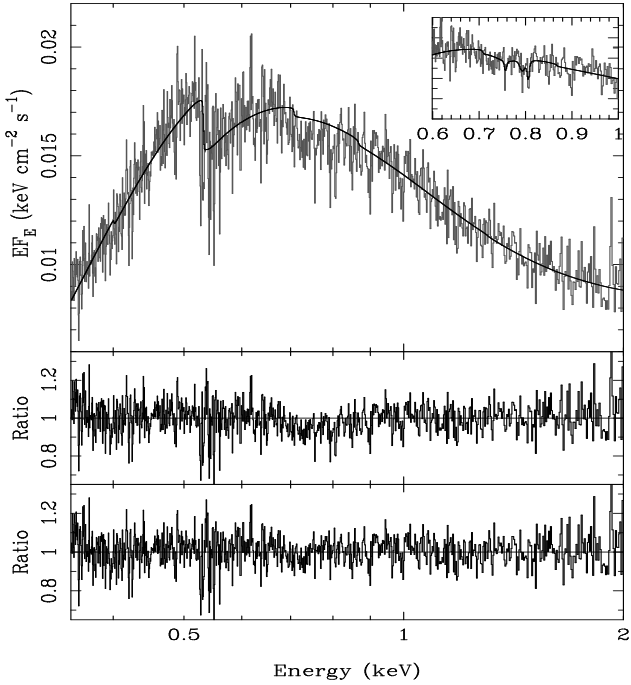} 
\caption[]{Top panel: The RGS spectrum of Ark~564 and the best fitting 
continuum model (black line) in the $<0.6$ and $>0.9$ keV band. The spectrum
shown is fluxed (i.e. the data points are divided by the effective area) but
not unfolded. Inset: close up of the 0.6-1 keV region with absorption lines
added to the model to represent the 0.7-0.9 keV Fe~UTA features. Middle panel:
data/model ratio for the continuum model fit in the same energy band. Bottom
panel: data/model ratio when absorption lines are included in the model to
represent the 0.7-0.9 keV Fe~UTA feature.}
\label{figure:RGS}
\end{figure}

The model fits well the overall shape of the RGS spectrum over the 0.35-2 keV 
range, except for some strong residuals around the Galactic {\footnotesize 
O\,I} edge (as with the EPIC spectra), a number of narrow absorption lines, and
a broad deficit of flux in  the 0.65-0.85 keV range. Interestingly, this 
residual feature is almost identical to a similar deficit that appears in the
EPIC residuals plot as well (Fig.\,\ref{figure:softpn} and
Fig.\,\ref{figure:softmos}).  The RGS spectrum, the continuum model (fitted to
the data below 0.6 keV and above 0.9 keV), and the data/model ratio are shown
in the top and middle panels of Fig.\,\ref{figure:RGS}, respectively. The
location of the broad dip centered at 0.77 keV corresponds to the blended
unresolved transition arrays of Fe\,I-XVI. A detailed study of this feature and
of the narrow absorption lines  in the Ark~564 RGS spectrum, using the
sophisticated warm absorber models available in the SPEX spectral fitting
package, will be presented elsewhere (Smith et al. in preparation).  For the
purposes of this work, in order to represent the absorption in the Fe~UTA
region, we added  4 Gaussian absorption lines to the continuum model and fitted
again the spectrum using XSPEC.

The results are shown in the inset panel of Fig.\,\ref{figure:RGS}, while the
data/model ratio is plotted in the bottom panel of the same Figure. The model
now fits well the overall shape of the entire spectrum, but the statistical
goodness of fit is still poor ($\chi^{2}_{\rm red}$/dof = 1.97/922), primarily
because of the structure around the Galactic {\footnotesize O\,I} edge, and
other narrow absorption features in the spectrum. Although the remaining
absorption lines and features in the RGS spectrum of \ark\ outside the Fe UTA
region are significant, they are too weak to warrant inclusion in the EPIC
spectral model. 

In a previous study of the X-ray spectrum of \ark\ using two short  XMM-Newton
observations,  Vignali \etal\ (2004) included a strong photoelectric edge from
{\footnotesize O\,VII} in their spectral model. Such an edge, which has a
threshold energy of 0.72 keV, is not visible in the RGS spectrum
(Fig.\,\ref{figure:RGS}). Replacing the UTA features in our spectral model with
an {\footnotesize O\,VII} edge yields $\chi^{2}_{red}$/dof of 2.02/927, a
poorer fit by $\Delta \chi^{2}=46$ with respect to our best fit model. The
fitted optical depth of the edge is $\tau=0.06\pm 0.02$, much smaller than the
$\tau=0.4\pm 0.1$ reported by Vignali \etal\ (2004),  but similar to the
optical depth determined by Matsumoto \etal\ (2004) in their HETGS observation 
($\tau=0.07^{+0.04}_{-0.05}$). If both the edge and the 4-Gaussian
representation of the Fe UTA features are included in the  spectral model  we
obtain $\chi^{2}_{red}$/dof of 1.97/921, no better than the model without the
edge. In this case the O\,VII edge is insignificant, with  $\tau<0.06$
($3\sigma$ limit).

We conclude that the soft band spectrum of \ark\ is indeed  intrinsically
smooth. The most significant spectral feature appears in the $0.7-0.9$ keV
band, and most probably corresponds to the blended unresolved transition arrays
of Fe\,I-XVI. In particular, we do not observe any strong absorption edges
around $0.7$ keV or any emission features around $0.9-1.1$ keV. 

%------------------------------ figure 11 ----------------------------------
 
%========================================================================
\subsubsection{The EPIC PN spectrum revisited}
%==========================================================================

%------------------------------ figure 12 ----------------------------------
\begin{figure}
\psfig{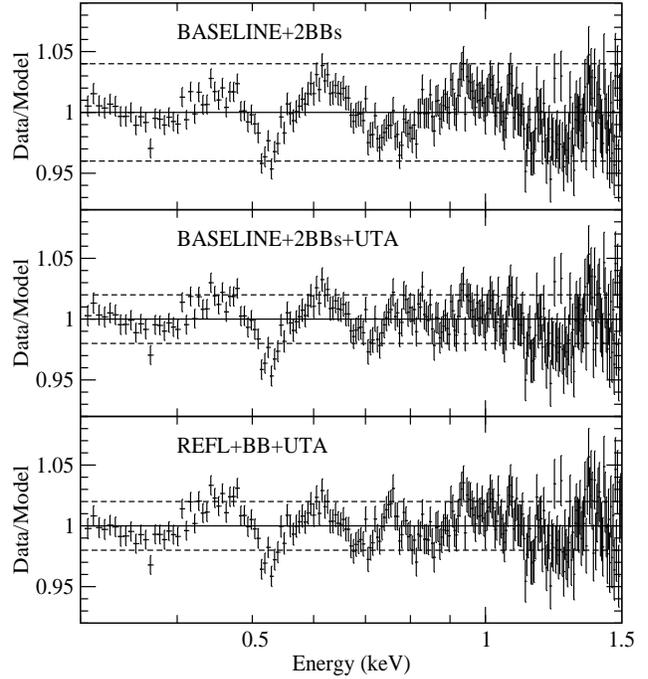}
 \caption[]{Data to model ratio of the ``{\footnotesize BASELINE+2\,BBs}"  (top
panel), the ``{\footnotesize BASELINE+2\,BBs+UTA}" (middle panel) and the
``{\footnotesize REFL+BB+UTA}" model fits to the  PN data in the $0.3-1.5$\,keV
band. For comparison reasons, the y-axis is identical in both panels.}
\label{figure:finalratio}
\end{figure}

In the top panel of Fig.\,\ref{figure:finalratio} we show the data/model ratio
in the case of the {\footnotesize BASELINE+2\,BBs} model fits to the PN
spectrum (for clarity reasons, in Fig.\,\ref{figure:finalratio} we show the
data/model ratio over the softest band only). As we mentioned earlier, the
deviations are at the $\la$ 4\% level (over the  whole energy band). Since the
results from the  RGS spectral analysis have  shown that the broad, shallow
deficit of flux in  the 0.65-0.85 keV range is most probably intrinsic, we
fitted again the EPIC PN spectrum with the {\footnotesize BASELINE+2\,BBs} and 
{\footnotesize BASELINE+BREMS} models, including the  4 Gaussian lines that we
used to model the absorption in the Fe~UTA region. Their centroid energy, width
and normalization  were kept fixed at the values we got from the RGS model
fitting. The addition of these 4 lines yields $\chi^{2}_{red}$/dof of 1.22/1171
and 1.36/1173, respectively. Compared to the simple  {\footnotesize
BASELINE+BREMS} model, the new model fit is poorer by $\Delta \chi^{2}=12.2$.
On the other hand, the addition of the  absorption lines representing the
Fe~UTA to the simple {\footnotesize BASELINE+2\,BBs} model,  improves the model
fit by $\Delta \chi^{2}=74.7$, with no additional degrees of freedom. 

This result demonstrates that the large $\chi^{2}_{red}$ values associated to
the models listed in Table 2 are due, to a large extend, to the presence of
warm absorber features. Indeed, in the middle panel of
Fig.\,\ref{figure:finalratio}  we show the data/model ratio in the case of  the
{\footnotesize BASELINE+2\,BBs+UTA} model. The model deviations are now reduced
to the $\la$ 2\% level over almost the entire energy band.  

However, the best fitting kT parameter values of the two black body components
are $\sim 0.15$ keV and $\sim 0.07$ keV. These temperatures are  rather high
and difficult to explain in terms of a simple accretion disc predictions (see
Sect.~5). For these reasons, we furthermore examined two alternative models
which are physically more motivated  and have already been used to model the
soft excess successfully in a few AGN.

The first one involves Comptonization. A two-temperature distribution of hot
electrons can in principle lead to the formation of both the hard power-law and
the soft excess.  In order to investigate this possibility we fitted the broad
band spectrum of \ark\ with  the {\footnotesize BASELINE}  plus a
{\footnotesize COMPTT} component (Titarchuk 1994) to account for the soft
excess, a black body component (to account for intrinsic emission from the
innermost disc) and the  4 narrow Gaussian lines for the UTA absorption
feature. This  model does not provide a good fit to the full band spectrum of
\ark\  ($\chi^{2}_{red}=1.89/1167$ dof), with large amplitude residuals
appearing in the soft band. 

An alternative possibility is that the soft excess emission is due to
reprocessing of the primary X-rays in the accretion disc. If the disc surface
is highly ionized, a significant excess of emission is expected to emerge  in
the soft X-ray band, including strong emission lines from the  irradiated disc.
If the emission features are smeared out by the motion of the disc, or by
general relativistic effects, then the soft excess emission could appear
smooth. In order to investigate this possibility, we used the
ionized-reflection model {\footnotesize REFLION} of Ross \& Fabian (2005) in
{\footnotesize XSPEC} together with a  power-law component (to account for the
illuminating source).  In order to take into account the Doppler and
gravitational effects around a black hole we adopted a simplified approach and 
convolved the emitted spectrum with a Gaussian of $v=0.2$c width. We also
added the narrow Gaussian line for the absorption feature at $8.1$ keV, a
black body component, and the four  narrow Gaussian absorption lines at $\sim
0.75-0.8$ keV (``{\footnotesize REFL+BB+UTA}" model).  

The model fits the broad band spectrum as well as the {\footnotesize
BASELINE+2\,BBs+UTA} model ($\chi^{2}_{red}=1.23/1169$ dof). In the bottom
panel of  Fig.\,\ref{figure:finalratio} we show the data/model ratio in this
case. The best fit model parameters are $\Gamma\sim 2.46$, kT$\sim 0.08$ keV
for the black body temperature, iron abundance $\sim$ solar, and ionization
parameter $\xi$\footnote {$\xi =L/nr^{2}$,  where $L$ is the hydrogen ionizing
luminosity, $n$ is the density, and $r$ is the distance between the absorber
and the source of ionization radiation.}$\sim  2000$ erg cm s$^{-1}$. The
measured (unabsorbed) ratio of the reflected flux to the total observed flux is
$\sim 0.45$.  

For both models, the largest amplitude residuals appear around the Galactic
{\footnotesize O\,I} edge. In order to investigate this issue further, we
repeated the model fits with the N$_{\rm H}$ and the oxygen abundance
parameters set free to vary during the fitting process. We found no significant
improvement to the previous model fitting results. Most probably, the residuals
around $\sim 0.5-0.6$ keV are caused by the current uncertainties associated
with the proper calibration of the instrumental and Galactic feature at 0.543
keV in the PN. 

With the Fe UTA features included in the PN fit, there is little evidence for
an emission line at around 1~keV. Nevertheless, in order to investigate this
issue further, we added a narrow Gaussian line component (with $\sigma$ fixed
at 5 eV) in the {\footnotesize BASELINE+2\,BBs+UTA} model and fitted again the
PN spectrum. The resulting line has a centroid energy of $\sim 0.96$ keV and an
equivalent width of $\sim 1$ eV. However, we get an improvement in the 
goodness of fit by $\Delta\chi^{2}=6.2$ for 2 additional degrees of freedom, 
which is not significant.

We conclude that, when we consider the UTA absorption features in the $0.7-0.8$
keV band, both the {\footnotesize BASELINE+2\,BBs} and {\footnotesize
REFL+BB+UTA} models fit well the full band X-ray spectrum of \ark. The fits are
not formally acceptable. However, the magnitude of the data/model residuals is
comparable to the accuracy with which the PN calibration is currently known.
This result, together with the ``smoothness" of the \ark/3C 273 PN spectrum 
(Fig.\,\ref{figure:ratio}), suggest that the statistically significant
discrepancies found between the data and the two models may well reflect, to a
large extent, remaining calibration uncertainties.

\section{Discussion}

{\it Spectral variability.} As we mentioned in Section 3, the large amplitude
flux variations of \ark\ are associated with spectral variations as well.
Changes in the continuum spectral shape can provide us with important clues
regarding the nature of the physical mechanism responsible for the X-ray
emission in the source. Furthermore, time-average spectral analysis  can miss
valuable information from the temporal behaviour of the emission/absorption
features in the spectrum. 

We plan to study in detail the spectral variability properties of the source in
the near future (Brinkmann \etal, in preparation). Nevertheless, in order to
give an idea of how the spectral characteristics of the source change with
time, we generated two {\footnotesize EPIC PN} spectra  (with a minimum of 30
counts/bin as for the time-averaged spectrum) for the time periods between
10-15 ks and 67-70 ks  after the start of the observation. As can be seen from
Fig.~\ref{figure:lc}, these spectra correspond to time periods when the
hardness ratio was low and high (we call them as the HRL and HRH spectra,
respectively). In both cases, the $3-11$ keV spectra are well fitted by a
simple {\footnotesize PL}  model ($\Gamma_{\rm HRL}=2.53\pm 0.11$, $\Gamma_{\rm
HRH}=2.25\pm 0.12$, and  $\chi^{2}_{red, \rm HRL}=0.96/103$ dof,
$\chi^{2}_{red, \rm HRH}=0.94/74$ dof). The full band spectra can be well
fitted by the {\footnotesize BASELINE+2BBs} model ($\chi^{2}_{\rm
HRL}=557.8/537$, and $\chi^{2}_{\rm HRH}=505.4/468$ dof).

We found no indication of any emission and/or absorption  features at $\sim
6.5-7$ keV and $\sim 8$ keV, and we could not detect the UTA feature in the
soft band, in either of the two spectra. This is not surprising, given the
small amplitude of these features in the time-average spectrum of the source.
Consequently, these features can be studied best with the use of the
time-average spectrum (which we present in this work), since, due to their
weakness, they can hardly be detected in spectra which are extracted from
shorter time periods.  

The best fitting kT values in the case of the HRL spectrum are $0.16\pm 0.01$
and $8.1\pm 0.5\times 10^{-2}$ keV. The respective values in the case of the
HRH spectrum are $0.17\pm 0.01$ and $7.7\pm 0.3\times 10^{-2}$ keV. Thus,
although the soft component's flux varied between the two periods, its shape
remained almost the same. In fact, Brinkmann \etal\ show that the shape of this
component remains roughly constant throughout the present \xmm\ observation.
Hence, the spectral shape of the soft excess component can also be studied best
with the time-average spectrum, as in this case we take advantage of the
highest possible signal-to-noise ratio data.

As for the hard band power-law component, the {\footnotesize PL} model fitting
to the HRL and HRH spectra suggest a spectral slope variation of the order of
$\Delta \Gamma \sim 0.3$, which is certainly not negligible.  The results from
a detailed study of the spectral slope variability will be presented in a
forthcoming paper by Brinkmann \etal\ For the time being, we can think of the
slope of  the time-average spectrum as a measure of the average spectral slope
during the \xmm\ observation of \ark.

{\it The hard band continuum.} A power-law model, with  $\Gamma = 2.43\pm0.03$
fits well both the PN and MOS time-average spectra at energies above 3 keV. 
This is in excellent agreement with the results from previous observations of
\ark. 

This  slope is quite steep even for a NLS1 (Leighly 1999b). It is interesting
to compare \ark\ with the Galactic black hole X-ray binaries (BHXBs) since  it
is widely believed that the physical processes that operate in AGN and BHXBs
are similar.  BHXBs exhibit photon indices steeper than 2.4 only when they are
in the so-called ``Very High" (VH) or  ``Steep Power-Law" (SPL) spectral  state
(McClintock \& Remillard, 2005). In this state, black hole binaries are usually
quite bright, with accretion rates higher than $\sim 0.2$ in Eddington units,
exactly like \ark\ which is also believed to be a high accretion rate system
(e.g. Romano \etal\ 2004). One possible explanation for the various BHXBs
spectral states invokes changes in the accretion flow geometry. For example,
Done (2002) argues in favour of the inner radius of the accretion disc 
decreasing with increasing accretion rate. At high accretion rates (like in
NLS1s), the disc could even extend down to the last stable orbit. At the same
time, Merloni (2003) argues that, if the  viscosity parameter is large,  high
accretion rates are  accompanied by powerful, magnetically dominated coronae.
Because of the changing disc geometry and the high accretion rate, the density
of soft photons is high, leading to a more efficient cooling of the corona.
Consequently, the resultant Comptonized spectrum from the corona becomes soft.
In other words,  the consistently steep hard band power-law in \ark\ implies
that the geometry of the accretion disc and/or the corona may be different to
that in the ``classical"  Seyfert galaxies. 

{\it The iron emission line.} We detect a significant iron line emission
feature at $\sim 6.7$ keV, in both the PN and MOS spectra. The line is weak
(EW$\sim$ 80 eV) and can  be well fitted with a narrow Gaussian line. It can
also be well fitted by a broad Gaussian line or a {\footnotesize DISKLINE}
model, but the use of these models is not statistically required. In fact, the
limited data quality in the iron line region due to its weakness prevents a
more precise determination of the line parameters. 

Our results are in agreement with those of Vaughan \etal\ (1999) who also
detected a rather week, i.e. EW$\sim 95$ eV, narrow, iron line in the \asca\
and \xte\ combined spectrum they studied. However, they are not  in agreement
with those of Turner \etal\ (2001) who detected a very strong (EW$\sim 650$ eV)
iron emission line during the one month long \asca\ monitoring in 2000, despite
the fact that the continuum slope and luminosity (above 3 keV) are comparable
in both cases. Therefore, variability in the line strength due to differences
in the primary radiation cannot account for the observed difference in the line
properties. The rather unusual magnitude of the previously reported feature,
and the low signal to noise ratio at the high energy end of a steep spectrum,
seem to point toward background subtraction problems in the earlier data. 

On the other hand, the line's energy is in agreement with the \asca\ results,
and suggests the presence of highly ionized material. The presence of a line
indicative of highly ionized iron is not surprising in the case of \ark, since
the incident X-ray spectrum is quite steep. In this case, the Compton
temperature even on the top of the layer responsible for the X-ray reflection
can be lower than $\sim 1$ keV. This allows some of the highest ionization
stages of iron to be abundant enough to imprint highly ionized signatures on
the X-ray spectrum, i.e. lines at 6.7 or 6.9 keV (Nayakshin, Kazanas \&
Kallman, 2000). 

{\it Warm absorbing material.}  The most  notable feature in the RGS spectrum
is a broad, shallow flux deficit in the energy  range $0.65-0.85$ keV, which
cannot be fitted well by an edge. Therefore, we do not confirm the presence of
the strong  photoelectric edge from {\footnotesize O\,VII} of $\tau\sim 0.4$
that Vignali \etal\ (2004) had reported from the spectral analysis of the
previous \xmm\ observations of \ark.  It is notable that Matsumoto \etal\ 
(2004) find that the {\footnotesize O\,VII} edge is very weak ($\tau=
0.07^{+0.04}_{-0.05}$) in their study of the HETGS spectrum of \ark, consistent
with our analysis. The study of the UV absorption lines by Crenshaw \etal
(2001) also predicts that the {\footnotesize O\,VII} edge should be small
($\tau\sim 0.035$), which is again consistent with the results from our
analysis.

It is possible that the report of a strong edge in Vignali \etal\ (2004) is due
to their use of a black body model to describe the soft excess in \ark, since 
we also observe a strong absorption  feature in the $0.6-0.75$ keV band  in the
best {\footnotesize BASELINE+BB} model fitting residuals (top panel in  
Figs.\,\ref{figure:softpn} and \ref{figure:softmos}). However, this  feature is
due to the incorrect form of the continuum model and does not correspond to an
intrinsic absorption feature in the spectrum of the source as the RGS data show
conclusively.

On the other hand, the broad flux deficit centered at 0.77 keV is similar to the
deep troughs that have been detected in the soft X-ray spectra of several
Seyfert galaxies and are inferred to be UTA of iron $n=2-3$ absorption lines
(e.g. Sako \etal\ 2001). The location and depth of the UTA implies it is
primarily due to material with an ionization parameter of $\xi\sim 10$, and a
column density of N$_{\rm Fe}\sim 10^{16}$\,cm$^{-2}$ (Behar, Sako, \& Kahn
2001), corresponding to N$_{\rm H}\sim 2\times 10^{20}$\,cm$^{-2}$ assuming
cosmic abundances.  The absorption features at the blue end of the UTA region
are due to Fe\,XVI and Fe\,XVII, and suggest additional material with $\xi\sim
100$, and N$_{\rm H}\sim 5\times 10^{20}$\,cm$^{-2}$.  This is broadly in
agreement with the estimate of Matsumoto \etal\ (2004), who also infer an
absorber with a range of ionization parameters from their {\it Chandra} HETGS
spectrum, although our estimate of the ionization parameter of the Fe~UTA
producing material is larger than theirs.

{\it The absorption feature at $\sim 8$ keV.} We find evidence  for an
absorption  feature in the PN spectrum  which can be well fitted by a narrow
Gaussian absorption line at $8.14\pm 0.04$ keV, with an equivalent width of 
$\sim 50$ eV. Absorption lines at energies $>6.5$ keV have been detected in the
spectra of a few nearby AGN, but with lower equivalent widths  (see Bianchi
\etal\ 2005). Recently, Risaliti \etal\ (2005) reported the detection of a
system of Fe\,XXV and Fe\,XXVI K$_\alpha$ and K$_\beta$ absorption lines in the
spectrum of NGC\,1365. Markowitz, Reeves, \& Braito (2006) also report the
detection of a narrow absorption line in the nearby Seyfert IC~4329a at 7.68
keV, which they attribute to Fe\,XXVI K$_\alpha$ absorption blueshifted to
$\sim 0.1$c relative to the systemic velocity. Additionally, absorption
features near $\sim 7-8$ keV, attributed to high-ionization Fe K-shell
absorption from material moving  at high velocities ($\sim 0.1-0.3$c) have been
detected in PG and Broad Absorption Line quasars, e.g. Reeves, O'Brien, \& Ward
(2003), Chartas \etal\ (2002), Chartas, Brandt, \& Gallagher (2003). 

Assuming a narrow line and an energy resolution of the PN detector of $\sim$
170\,eV at 8\,keV (Ehle et al. 2005) we can estimate the significance of this
feature from the counting statistics in the PN spectrum to be $\sim 4\,\sigma$.
However, if the feature we detect is indeed real, then it most probably
corresponds to iron absorption from material outflowing with high  velocity
(see below). In this case, we are in effect fitting the extra  absorption line
component over many energy bins, where narrow features may occur by chance
(i.e. statistical noise). A more conservative estimate of the significance of
the line detection can be made if we consider the number of spectral bins over
the energy range where one may expect to detect iron absorption lines (Porquet
\etal\ 2004).  In our case, we define this energy range to be from 6.7 keV (the
Fe\,{\footnotesize XXV} K$\alpha$ energy) up to 9.06 keV  (the energy of the 
Fe\,{\footnotesize XXVI} K$\alpha$ line from material outflowing at 0.3c).
There are 81 spectral bins (at 30 counts per bin) in the PN spectrum in this
energy range. Following Porquet \etal\ (2004), we estimate that the probability
of detecting the  absorption feature at 8.1 keV by chance is only  $\sim
2.6\times 10^{-3}$. 

We also carried out a more rigorous test of the significance of the line's
detection using Monte Carlo simulations like those performed by Porquet \etal\
(2004). We used the {\footnotesize XSPEC FAKE} command to create 1000 synthetic
PN spectra corresponding to the {\footnotesize PL+NLG} best-fitting model, with
photon statistics expected from a 100 ks exposure, and grouped to minimum 30
counts per bin. We fitted each synthetic spectrum with a  {\footnotesize
PL+NLG} model, and recorded the minimum $\chi^{2}_{\rm PL+NLG}$ value. Then we
added a narrow absorption line ($\sigma=0.01$ keV) to the model fit, with the
line energy restricted in the range 6.7$-$9.1 keV. In fact, we stepped the line
energy  over this energy range, in steps of size 0.1 keV, and fitted  the model
each time to ensure the lowest $\chi^{2}_{\rm +line}$ value was found. We then
obtained $\Delta\chi^{2}=\chi^{2}_{\rm PL+NLG} - \chi^{2}_{\rm +line}$ for each
synthetic spectrum. Using these 1000 synthetic values we constructed the
sample  cumulative distribution function of $\Delta\chi^{2}$ in the case of a
``blind search" in the 6.7$-$9.1 keV range, under the assumption there is no
real absorption line there. In the case of the \ark\ PN spectrum, the addition
of a narrow absorption line to the  {\footnotesize PL+NLG} model results in a
$\Delta\chi^{2}=18.97$. Such a large decrease in $\chi^{2}$ occured in just $6$
of the 1000 synthetic spectra. According to this result, the detection of the
absorption line at $\sim 8.1$ keV is significant at the 99.4\% confidence
level.

We conclude that it is indeed  possible that the absorption feature we detect
at 8.1 keV is real. This result implies the presence of a second warm absorbing
layer, with properties quite different to those mentioned in the previous
paragraph.  The most plausible assumption is that the line corresponds to 
Fe\,{\footnotesize XXVI} K$_\alpha$ at rest energy of 6.97\,keV. One would also
expect to detect the other highly ionized absorption lines at:
Fe\,{\footnotesize XXV}  K$_\alpha \sim 7.64$\,keV,  Fe\,{\footnotesize XXV}
K$_\beta\sim 8.99$\,keV and Fe\,{\footnotesize XXVI}  K$_\beta \sim 9.4$\,keV.
There are indeed flux depressions near these energies, but they are  not
significant and the line associations are marginal.  The lack of an observable 
Fe\,{\footnotesize XXV} K$_\alpha$ line implies highly ionized material while
the  EW of the line we observe implies a high absorbing column of N$_{\rm H}>
10^{23}$ cm$^{-2}$ (Risaliti \etal\ 2005). Under these conditions, the EW of
the  Fe\,{\footnotesize XXVI} K$_\beta$ line will be  2.5 times smaller than
that of Fe\,XXVI K$_\alpha$, and this can explain the lack of its detection. 

However, if the feature we detect is indeed the  Fe\,{\footnotesize XXVI}
K$_\alpha$ line, then it is blue-shifted in the source by (1+z)$\sim$1.17 (i.e.
outflow velocities of 50000 km/s=0.17$c$). Note that this velocity is not
consistent with the recession velocity of \ark, which is $\sim 7400$ km/s.
Consequently, the absorption feature cannot be due to hot Galactic gas, as was
recently suggested for some of the absorption signatures  in the X-ray spectra
of a few AGN (McKernan, Yaqoob, \& Reynolds,  2005). 

The narrowness of the absorption feature is consistent with the presence of
just a single component, i.e. of a single ``blob" of material, ejected
presumably from the nucleus, as opposed to a continuous flow, which is what one
would  probably expect, i.e. a stream of gas with a large range of velocities,
indicative of its acceleration up to  velocities of $\sim 0.1-0.2$c. In any
case,  the absorption feature implies that a very effective acceleration
mechanism is in operation in \ark.  If the bolometric luminosity of the source 
is indeed comparable to (or even larger than) its Eddington luminosity, perhaps
Thomson scattering could provide the necessary acceleration to the warm
material. 

We finally note that  Vaughan \etal\ (1993) detected an absorption edge of
$\tau\sim0.2$ at 8.6 keV in their combined \asca\ and \xte\ spectrum of the
source. This feature is indicative of the presence of helium-like iron. The
authors considered the possibility of absorbing material which lies along the
line of sight and found that, in this case, a high ionisation parameter and
densities in excess of $10^{23}$ cm$^{-2}$ were needed to explain their
results. Although the feature we observe in the \xmm\ spectrum is  certainly
not an edge, we believe it is interesting that the properties of the absorbing
material are similar in both cases.

{\it The $\sim 1$ keV emission line.}   We do not observe any  significant
emission line features in the $0.9-1.1$ keV band either in the RGS or in the
EPIC best model fitting residuals plot (Figs.\,\ref{figure:softpn},
\ref{figure:softmos}, and \ref{figure:RGS}). The addition of such a component
to the PN spectral model does not result in a significant improvement in the
goodness  of fit.  We conclude that there is no strong indication of a
significant intrinsic  emission line at $\sim 0.9-1$ keV in the EPIC (and RGS)
data of \ark. Our results are in agreement with those of Matsumoto \etal\
(2004) who could not detect a prominent narrow emission line around 1 keV in
their {\it Chandra} HETGS spectrum. They  also ruled out the  possibility that
the 1 keV feature originates from blends of several narrow  emission lines. We
believe that, just like with the {\footnotesize O\,VII} edge, the  previous
claims of  excess emission at around 1\,keV (e.g.  Brandt \etal\ 1994, Turner,
George \& Netzer 1999, Comastri \etal\ 2001) might have been influenced by the
choice of a black body or a power-law model  to  reproduce the low energy
continuum.  We found no prominent narrow emission line around 1 keV and ruled
out the  possibility that the 1 keV feature originates from blends of several
narrow  emission lines.

{\it The soft excess shape.}   A solid result of our study is that the soft
excess cannot be parametrized  either by a multiple power-law or a power-law
plus a single black-body  model. In this respect, \ark\ is different  to other
NLS1s like for example Ton S180 (Vaughan \etal\ 2002) and  Mrk 478 (Marshall
\etal\ 2003), which show featureless power-law-like soft excess  components.
Furthermore, the soft excess emission is probably not resulting from thermal
Comptonization either.``Two-temperature" thermal Comptonization models have
been used in the past to fit successfully the broad band X-ray spectra of a few
NLS1s like e.g. Mrk 896 (Page \etal\ 2003) and PKS 0558-504 (Brinkmann \etal\
2004). This is not the case with \ark.

We find that, when we add the 4 Gaussian absorption lines to parametrize the
UTA feature, the best description of the soft excess shape in the PN data  is
provided by two black bodies (with kT$\sim 0.15$ and $\sim 0.07$ keV),  or a
black-body plus reflection model. The best model fitting residuals in the soft
band reduce to the 2\% level, which is comparable to the accuracy with which
the instrumental EPIC calibration is currently known. A two black body model
fits well the soft excess emission in other NLS1 galaxies as well (e.g.  Mrk
359, O'Brien \etal\ 2001).

However, the temperature of the kT$\sim 0.15$ keV black body component  is not
consistent with that expected from a simple accretion disc. If we assume that
the black hole mass in \ark\ is  $2.6\times 10^{6}$ M$_{\odot}$ (Botte \etal\
2004), then a standard accretion disc around it should have a peak temperature
$\la 20$ eV, even if we accept that the system is accreting at its Eddington
limit. This is not consistent even with the temperature of the cooler (i.e. the
kT$\sim 0.07$ keV) black body component in our model. Clearly the results from
the  2 BBs model is at odds with the expectations for a simple accretion disc.
This result is in agreement with Gierlinsky \& Done (2004) and Piconcelli
\etal\ (2005), who present  a detailed discussion on the problems arising when
the soft excess emission in AGN is identified as direct thermal emission from
the innermost part of the accretion disc.

A photoionized disc reflection model also fits the overall spectrum well. Our
best fit model implies that the iron abundance is consistent with being solar.
The measured fraction of reflected over the total observed flux is $\sim 0.45$,
close to 0.5 which is expected in the case of gas illuminated by an isotropic
source. Interestingly, this model fits well  the hard band spectrum as well.
Since it  reproduces the emission expected from a photoionized accretion disc
around a black hole, it includes the iron emission lines at $\sim 6.4-7$ keV.
The fact that the model is  consistent with the weak line in the \ark\ spectrum
is mainly due to the fact that it is convolved with a Gaussian in order to 
simulate the relativistic effects from a disc. 

Recently, Crummy \etal\ (2006) have shown that a relativistically blurred
photoionized disc reflection model can fit well the broad band X-ray spectra of
a large sample of type 1 AGN. One of the objects in their sample is \ark. They
also find a good agreement between the model and the $0.3-11$ keV spectrum of
the source (using data from a past \xmm\ observation).  Their best fit model
parameters are somehow different to ours ($\xi\sim 3100$, iron abundance over
solar $\sim 0.5$). They also detect an absorption edge of $\tau\sim 0.1$ at
$\sim 0.68$ keV. Although this is not as large as the one reported by Vignali
\etal\ (2004), a $\tau\sim 0.1$ edge  would still be detected easily in the RGS
spectrum. If we take out the black body component from our model, we also
observe a strong edge-like feature in the residuals. Since such a strong edge
is probably not intrinsic, as we have mentioned before, we believe that the use
of a separate black body component, instead of an absorption edge, is a more
realistic approach to the modeling of the soft excess in \ark. On the other
hand, the best fit temperature of the black body component is still quite high
($\sim 0.08$ keV). Further work, with the use of models more sophisticated than
a simple black body, is needed in order to investigate this issue further. 

\section{Conclusions}

We have presented the results from the spectral analysis of a $\sim$ 100\,ksec
observation of \ark\ by \xmm\ which yielded data with unprecedented quality for
this source. Our main aim in this work was the study of the time-average
$0.3-11$ keV spectrum, using the EPIC PN and MOS data. Using the RGS data, we
also investigated the presence of any significant absorption/emission features
that could  affect the EPIC data. Our results from the detailed study of the
RGS spectrum, the timing properties of the source, and from time resolved
spectral analysis will  be presented in forthcoming papers. 

We find that both the PN and MOS spectra above 3 keV are  well fitted by a
steep power-law of $\Gamma=2.43$. 

We observe an iron emission line at $\sim 6.7$ keV, indicative of the presence
of ionized material in the vicinity of the central source. The line is week
(EW$\sim 80$ eV) and narrow. Although the possibility of a broad line is
consistent with the data, it is certainly not required.

At energies below 2 keV, both the PN and MOS  spectra steepen, and are
dominated by a broad soft excess emission. When compared to the hard X-ray
power-law, the soft excess rise flattens below $\sim 0.6$ keV. 

All our attempts to fit the soft excess were formally not acceptable. However,
a model that consists of two black bodies or a black body and a reflection
component do fit the soft band EPIC PN spectrum reasonably well, with the
amplitude of the remaining model residuals being less than 2\%. To a large
degree, the fact that these model fits are not statistically acceptable  is due
to the exceedingly high statistical quality of the data. As a result, at the
lowest energies, the model fits are certainly affected by remaining calibration
uncertainties. More accurate knowledge of the EPIC PN calibration is needed in
order to constrain meaningfully the best fit model parameters in \ark.

When compared with the simple accretion disc predictions, the temperature of 
the black body components in the {\footnotesize BASELINE+2BBs} model are
unreasonably high. For that reason, we believe that the {\footnotesize REFL+BB}
model provides a more physically justified description of the full band X-ray
spectrum of \ark. Our results suggest the presence of highly ionized material
(not surprising, given the high X-ray luminosity of the source) in a
geometrically flat disc  illuminated by an isotropic source, with solar
abundances. The weakness of the iron emission line could be explained by
relativistic effects which tend to smooth out the reflected component. 

Finally, we do not confirm the presence of the {\footnotesize OVII} edge and an
emission line at $\sim 1$ keV. Instead, we detect a broad, shallow flux deficit
in both the EPIC and RGS spectra which corresponds to the blended unresolved
transition arrays of Fe {\footnotesize I-XVI}, and, for the first time in the
case of \ark, an absorption line at $\sim 8.1$ keV in the PN spectrum. The
location and depth of the UTA region implies the presence of warm absorbing
material with $N_{\rm H}\sim (2-5)\times 10^{20}$ cm$^{-2}$. If the absorption
line corresponds to Fe {\footnotesize XXVI} K$\alpha$, it suggests the presence
of highly ionized,  absorbing material of $N_{\rm H}> 10^{23}$  cm$^{-2}$,
which is moving away from the central source at a high velocity of $\sim
0.17c$.

\vskip 0.4cm
\begin{acknowledgements}
This work is based on observations with \xmm, an ESA science mission with
instruments and contributions directly funded by ESA Member States and the USA
(NASA). We gratefully acknowledge travel support through the bilateral
Greek-German IKYDA2004 personnel exchange program. PU acknowledges support from
an EU Marie Curie Fellowship.
\end{acknowledgements}

\end{document}